 \documentclass[11pt,a4paper]{article}

\usepackage[english]{babel}
\usepackage{a4}
\usepackage{a4wide}
\usepackage[numbers]{natbib}
\usepackage{dsfont}
\usepackage{graphicx}
\usepackage{psfrag}
\usepackage{amsmath}
\usepackage{amssymb}
\usepackage{slashed}
\usepackage{verbatim}

\newcommand{\be}{\begin{equation}}
\newcommand{\ee}{\end{equation}}
\newcommand{\bea}{\begin{eqnarray}}
\newcommand{\eea}{\end{eqnarray}}
\newcommand{\bean}{\begin{eqnarray*}}
\newcommand{\eean}{\end{eqnarray*}}

\newcommand{\irm}{{\rm i}}

\newcommand{\cl}[1]{{\mathcal #1}}

\newcommand{\pa}{\partial}

\newcommand{\ds}{\displaystyle}

\newcommand{\bC}{\mathds{C}}

\newcommand{\clL}{\cl{L}}

\newcommand{\eq}[1]{(\ref{#1})}

\newcommand{\tr}{\mathrm{tr}\!}

\renewcommand{\twocolumn}[2]
{\left(\begin{array}{c} #1 \\ #2 \end{array}\right)}
\newcommand{\twomatrix}[4]
{\left(\begin{array}{cc}#1 & #2 \\ #3 & #4\end{array}\right)}

\newcommand{\qed}{\nobreak \ifvmode \relax \else
      \ifdim\lastskip<1.5em \hskip-\lastskip
      \hskip1.5em plus0em minus0.5em \fi \nobreak
      \vrule height0.75em width0.5em depth0.25em\fi}

\newcommand{\SL}{\mathrm{SL(2,\bC)}}
\newcommand{\SO}{\mathrm{SO(1,3)}}
\newcommand{\SU}{\mathrm{SU(2)}}

\renewcommand{\sl}{\mathrm{sl(2,\bC)}}
\newcommand{\so}{\mathrm{so(1,3)}}
\newcommand{\Wb}{\overline{W}}

\newcommand{\psib}{\overline{\psi}}

\newcommand{\nub}{\overline{\nu}}
\newcommand{\eb}{\overline{e}}
\newcommand{\sigmab}{\overline{\sigma}}

\newcommand{\Lambdab}{\overline{\Lambda}}
\newcommand{\Ab}{\overline{A}}
\newcommand{\Fb}{\overline{F}}
\newcommand{\ad}{\dot{a}}
\newcommand{\bd}{\dot{b}}
\newcommand{\cd}{\dot{c}}
\newcommand{\dd}{\dot{d}}
\newcommand{\At}{\tilde{A}}
\newcommand{\Ft}{\tilde{F}}
\newcommand{\Dt}{\tilde{D}}

\begin{document}

\renewcommand{\thefootnote}{\fnsymbol{footnote}}
\begin{center}
{\bfseries\LARGE Isogravity: Toward an Electroweak and Gravitational Unification} \\[4mm]
\large Stephon Alexander\footnotemark[1] \\[3mm] 
{\large\itshape Institute for Gravitational Physics and Geometry, \\ Physics Department, Penn State University, \\ University Park, PA 16802, U.S.A}
\end{center}
\footnotetext[1]{email: \tt alexander at gravity dot psu dot edu}
\footnotetext[7]{email: \tt conrady at gravity dot psu dot edu}
\renewcommand{\thefootnote}{\arabic{footnote}}
\setcounter{footnote}{0}
\vskip.5cm

\begin{abstract}

We present a model that unites the electroweak interaction with general relativity.  This is made possible by embedding the kinetic terms for gravity and electroweak theory using one $\SL$ connection variable.  The gauge theory is specified without relying on a space-time metric.  We show that once a symmetry breaking mechanism is implemented that selects a global time-like direction, the elecroweak theory and general relativity emerges with their associated massless degrees of freedom; the spin 1 vector boson and the spin 2 graviton.   

\noindent 
\end{abstract}
\vskip1cm

\setlength{\jot}{0.3cm}

\section{Introduction}

In this paper, we propose a new path toward unifying the electroweak standard model with general relativity. Despite the great successes of the standard model, it is well known that it does not incorporate gravity.  The key conceptual leap which clears a way toward this path, is the ability to express a $\SU $gauge theory encoding both the electroweak interactions and gravity as a chiral gauge theory without resort to a background space-time metric (ie. apriori, there is no distinction between internal and external 'space').  Much of this insight is based on using the wisdom of the chiral, self-dual Ashtekar variables\cite{Ashtekar:1986yd}.  But why should one single out gravity and the electroweak inteactions for unification?  First, the Electroweak theory is a chiral theory which maximally violates parity. Likewise, gravity formulated as a gauge theory a la. the Ashtekar self-dual variable is a chiral gauge theory.  

The electroweak interaction and gravity distinguish themselves from the other interactions in that they both interact universally with quarks and leptons.  The difference between these two interactions arises when one considers fermionic matter; the gravitational connection acts on the spin angular momentum of the fermions while the electroweak connection acts on the isospin degree of freedom in a parity violating manner.   Is it a coincidence that the groups associated with both isospin and spinorial angular momentum is $SU(2)$? Can one imagine a new symmetry which relates these roles of gravity and the electroweak theory?  In this paper we show that both theories can arise from an underlying gauge theory which unifies general relativity and electroweak interactions.  Our proposal is also based on the breaking of $\SL$ gauge symmetry to an $\SU$ gauge symmetry according to Dell, de Lyra and Smolin\cite{cahill,Dell:1986pw}.

So that we can combine these two ideas, we start with the $\SL$ group which transforms the left-handed.
The pure gauge part of the action contains the action of Ashtekar gravity for the left and right connection, as well as Yang-Mills-like terms for them. The Yang-Mills terms are stabilized by the presence of two copies of Hermitian inner products, analogous to the inner product in the model of Dell, de Lyra and Smolin.

In the matter sector, we have left handed Fermions, which is endowed with \textit{two} 2-spinor indices. One of these two indices is contracted with the $\sigma$ of the Weyl action, while the other index is contracted with the inner product. Due to this choice, the first index will take on the conventional meaning of a spinorial index, while the second index will be interpreted as an $\SU$ color index (see below).

When we consider the classical equations of motion of the unified gauge theory, we obtain the zero torsion condition for the left-handed connection, so they are forced to be the self-dual part of a single connection which is compatible with the tetrad. In this way, we recover Einstein gravity classically.  By consistently solving the background field equations and their fluctuations, for the theory in the broken phase, we find the astonishing result that the chiral connection plays a dual role as a transverse traceless graviton and the W and Z boson (although the vector bosons are massless in this theory).   

Moreover, we find a connection between parity violation and chirality. Namely, the chiral structure of gravity which is encoded in Ashtekar's formulation of gravity is reflected in the chiral interactions in the electroweak theory.   The resulting gauge theory is left non invariant under parity reflection since the underlying $Z_{2}$ symmetry is broken in the Isogravity Lagrangian.

At the quantum level, the left handed part of $\SL$ is dynamically broken to an $\SU$'s by the background value of a gauge fixing . As a result, we get massive timelike component of the vector boson that correspond to the quotient $\SL/\SU$. They can be chosen sufficiently massive, so as to be unobservable at present. 

In section I we motivate and derive the theory which unifies gravity and the electroweak interaction. In section II, we discuss the inclusion of chiral Fermions.  In section III, we demonstrate how classical general relativity and the electroweak theory emerges from the unified Lagrangian due to the gauge fixing procedure.  We then discuss, in Section V, how both the graviton and the massless vector boson arises from the vacuum solution of the master theory.  In Section VI, we derive the standard and non standard interactions in the Leptonic sector of the electroweak theory.   

\section{Isogravity: Field Content and Lagrangian}

\subsection{Gauge sector}

It is well known that general relativity can be formulated as a gauge theory without resorting to the space-time metric.  It was realized that working with the chiral, self-dual part of a complex $\SL$ connection is equivalent to real general relativity.  Our first step toward gravity-weak unification is to simply add a kinetic curvature term to the chiral gravity theory with one chiral gauge connection. Schematically we will be dealing with an action of the form:

\be S= \int_{M_{4}} R(E,A) + R \wedge *R \ee
where R(E,A) is the curvature of the gauge connection.  But why should we expect this relatively simple theory with just one connection to encode two forces?  

The key is to use the  gauge group:

\be
SO(3,1;C) = \SL_L \times \SL_R 
\ee
where $\SL_L$ and $\SL_R$ are two copies of $\SL$: the ``left-handed'' group $\SL_L$ is formed by the special linear maps on a complex 2-dimensional vector space $W$, while $\SL_R$ consists of the special linear maps on the dual complex conjugate space $\Wb^*$. We denote the associated connections by $A^L$ and $A^R$. 

The generators of the left and right part of group act independently and are related to each other by a discrete parity transformations on the complex spinors(ie. the representations are not unitarily equivalent). We will now write down a gauge theory, where the electroweak isospin symmetry is generated by the left handed part of the complexified Lorentz group.  Likewise, the gravitational curvature is defined solely in terms of the left connection.   

The curvature is given by
\bea
F^L_{\mu\nu} &=& \pa_\mu A^L_\nu - \pa_\nu A^L_\mu + [A^L_\mu,A^L_\nu]\,, \\
\eea
We denote representations of the left handed sector of the gauge group by $(k,l)$, where the first two indices correspond to the usual classification of spinors.

In addition to the connections, we have four real spinor fields $\sigma^\mu$, $\mu = 0,1,2,3$, of type $(0,1)$:
\be
\sigma^\mu{}_{a\bd} = \overline{\sigma}^\mu{}_{b\ad}
\ee
At each point, they can be decomposed into four linearly independent components $\sigma^I$,
\be
\sigma^\mu = E_I{}^\mu \sigma^I\,, 
\ee
where\footnote{Spinor indices are raised and lowered by $\epsilon$-tensors according to the standard conventions.}
\[
\renewcommand{\arraystretch}{2}
\begin{array}{l@{\qquad}l@{\quad}l}
(\sigma^0{}_{a\ad}) = \mathds{1}\,, & (\sigma^i{}_{a\ad})^\mathrm{T} = -\sigma^i{}\,, & \\
(\sigma^0{}^{a\ad}) = \mathds{1}\,, & (\sigma^i{}^{a\ad}) = \sigma^i\,, & i = 1,2,3\,.
\end{array}
\]
The coefficients $E_I{}^\mu$ are called tetrads, and we assume that $\det(E_I{}^\mu) \neq 0$. Furthermore, we have a hermitian, positive definite spinor field$s$ of determinant one and type $(0,1)$.  This field plays the role as the metric on the internal isospin space.  It is comprised a matrix of scalar fields which dynamically reduces the weak isospin gauge group from $\SL$ to $SU(2)$ by a gauge fixing.   Before this gauge fixing, the theory is $Z_{2}$ symmetric in interchange the weak isospin and spinor rotations.

We refer to $s_L$ as the left inner product which will only act on the isospin index.

For the pure gauge sector, we will show that following Lagrangian density encodes the Electroweak interactions and general relativity: 
\bea
\clL_{\mathrm{gauge}} &=&
\det(E)\left[         
\frac{1}{16\pi G}\left(
E_I{}^\mu\,\sigma^I{}_{a\ad}\,E_J{}^\nu\,\sigma^J{}^{b\ad}\,F^L_{\mu\nu}{}^a{}_b + \mathrm{c.c.}  
\right)\right. \nonumber \\
&& 
\hspace{1.3cm}{}- \frac{1}{4 g^2}\left(
(s^{-1})^{d \ad}\,\Fb^L_{\mu\nu}{}^{\bd}{}_{\ad}\,s{}_{c \bd}\,F^{L\mu\nu}{}^c{}_d \right) \\
&& \hspace{1.3cm}{}+  \frac{m^2}{16}\,\left((s^{-1})^{a \dd} D_\mu s{}_{c\dd}\right)\left((s^{-1})^L{}^{c \bd} D^\mu s{}_{a\bd}\right) 
\eea

Here, $\det(E)$ stands for the determinant of the co-tetrads
\be
\det(E) \equiv \left(\det(E_I^{-1}{}^\mu)\right)\,,
\ee
and the spinor $(s^{-1})^{a \bd}$ are defined by 
\be
(s^{-1})^{a \bd}\, s{}_{c\bd} = \delta^a{}_c\,,\ee 
The covariant derivative on $s$ is 
\bea
D_\mu s{}_{a\bd} &=& \pa_\mu s{}_{a\bd} - s{}_{c\bd}\,A^L_\mu{}^c{}_a - \Ab^L_\mu{}^{\dd}{}_{\bd}\,s{}_{a\dd}.
\eea

We can write the Lagrangian more compactly by introducing matrix notation: after setting 
\be
\sigmab^I = (\sigma^I{}_{a\ad})^\mathrm{T}\,,\qquad
\sigma^I = (\sigma^I{}^{a\ad})\,,
\ee
and
\be
s = \left(s{}_{a\bd}\right)^\mathrm{T}\,,\qquad 
s^{-1} = \left(\left(s^{-1}\right){}^{a\bd}\right)\,,
\ee
the Bosonic sector of the Lagrangian can be written as 
\bea
\clL_{\mathrm{gauge}} &=& 
\det(E)\left[
\frac{1}{16\pi G}\,E_I{}^\mu E_J{}^\nu\left(\tr\left[\sigma^I\,\sigmab^J F^L_{\mu\nu}\right] + \mathrm{h.c.} \right) \right. \\
&& \hspace{1.3cm}{}- \frac{1}{4 g^2}\left(\tr\left[s^{-1}\,F^{L\dagger}_{\mu\nu}\,s\,F^L{}^{\mu\nu}\right] \right)\\
&& \hspace{1.3cm}{}+ \frac{m^2}{16}\,\tr\left[\left(s^{-1} D_\mu s\right)\left(s^{-1} D^\mu s\right)\right] \
\label{gaugesector}
\eea
The equation 
\be
\Lambda^I{}_J = \sigma^I{}_{a\ad}\,\sigma_J{}^{b\bd}\,\Lambda^a{}_b\, \Lambdab^{\ad}{}_{\bd}
\ee
determines the homomorphism  between $\SO$-- and $\SL$--transformations. 

We see from this that the Lagrangian density \eq{gaugesector} is invariant under a local Lorentz transformation that transforms 
connection, field strength, tetrads and inner products as follows:
\be
\renewcommand{\arraystretch}{1.7}
\begin{array}{lcl}
A^L &\to& \Lambda\,A^L\,\Lambda^{-1} + \pa\Lambda\,\Lambda^{-1}\,, \\
F^L &\to& \Lambda\,F^L\,\Lambda^{-1}\,, \\

E_I{}^\mu &\to&  E_J{}^\mu\,(\Lambda^{-1})^J{}_I\,, \\
s &\to& \Lambda^\dagger{}^{-1}\,s\,\Lambda^{-1}\,, \\
\end{array}
\label{gaugetransformation}
\ee

\section{Fermionic Sector}

In this section we will demonstrate how chiral interactions with Fermions arise naturally in the Isogravity theory.  This happens because, as stated above, the isospin and the chiral connection transform in the same representation.  In formulating our gauge theory we were forced to choose one connection

For the fermionic content of the theory, we take a fermion field in the $(2,0)$ representation (called $\psi_L$). It will become evident below why the fermionic statistics is consistent with this representation assignment. The fermion coupling is chosen as
\be
\clL_{\mathrm{fermion}} = 
\det(E)\left(\irm\,\psib_L{}^{\ad\bd}\, E_I{}^\mu\,\sigma^I{}_{a\ad}\,s{}_{b\bd}\,D_\mu\psi_L{}^{ab} + \mathrm{h.c.} \right)\\
\label{fermionsector}
\ee
with the covariant derivative given by 
\bea
D_\mu\psi_L{}^{ab} &=& \pa_\mu\psi_L{}^{ab}  + A^L_\mu{}^a{}_c\,\psi_L{}^{cb} +  A^L_\mu{}^b{}_d\,\psi_L{}^{ad}  \\ 
\eea
We can translate \eq{fermionsector} to matrix notation as
\bea
\clL_{\mathrm{fermion}} = \det(E)\left(\irm\,\psi_L^\dagger\, E_I{}^\mu\,\sigmab^I\,s\,D_\mu\psi_L{} + \mathrm{h.c.} \right)\\
\eea
it is important to keep in mind that this notation somewhat does not explicitly indicate that the $\sigma$'s only contract with the first index of the $\psi$'s, while $s$ contracts only with the second index of the $\psi$'s.

So that the total Lagrangian is invariant under gauge transformations \eq{gaugetransformation}, the fermions have to transform as
\be
\renewcommand{\arraystretch}{1.7}
\begin{array}{lcl}
\psi_L{}^{ab} &\to& \Lambda^a{}_c\,\Lambda^b{}_d\,\psi_L{}^{cd}\,, \\
\phi_L{}^a &\to& \Lambda^a{}_c\,\phi_L{}^c\,, \\
\psi_R{}_{\ad \bd} &\to& (\Lambda^{-1})^{\cd}{}_{\ad}\,(\Lambda^{-1})^{\dd}{}_{\bd}\,\psi_R{}_{\cd \dd}\,, \\
\phi_R{}_{\ad} &\to& (\Lambda^{-1})^{\cd}{}_{\ad}\,\phi_R{}_{\cd}\,.
\end{array}
\label{gaugetransformationfermionsHiggs}
\ee

\section{Emergence of Einstein gravity and the electroweak model}

\subsection{Classical background}

In the previous section, we showed that the gauge fixing conditon on the internal metric, breaks the symmetry leading to a chiral and parity violating gauge Lagrangian which resembles the massless $\SU$ Electroweak theory.  The other part of the theory resembles General relativity.  We need to study the dynamics of the theory in the broken phase at the background classical level and at the level of linear perturbation theory. Moreover, we will need to show that our theory does indeed have a massless graviton.

We will now determine a classical solution of our action. In the next subsection, this classical solution will be taken as the background when we quantize the gauge theory.

To determine the background, we proceed in a stepwise fashion. At first, we will just consider the first two terms in the gauge Lagrangian: in that case, the Lagrangian is equivalent to the Einstein-Hilbert Lagrangian, and we can pick Minkowski spacetime as a solution. Then, we will add the other terms of the total Lagrangian, and see what consequences this has and if Minkowski spacetime is still a solution.

Let us start by showing that the two terms
\be
\clL_{\mathrm{gravity}} = 
\frac{1}{16\pi G}\,\det(E)\,E_I{}^\mu E_J{}^\nu
\left(\tr\left[\sigma^I\,\sigmab^J F^L_{\mu\nu}\right] + \mathrm{h.c.} \right) \\
\label{onlyfirsttwoterms}
\ee
are classically equivalent to the Einstein-Hilbert Lagrangian of gravity.

To prove this, note that the homomorphism
\be
\Lambda^I{}_J = \tr\left[\sigmab^I\,\Lambda^L\,\sigma_J\,\Lambda^L{}^\dagger\right]
\ee
between group elements of $\SL_L$ and $\SO$ leads to an isomorphism
\be
\omega^{IJ} = \tr\,\Big[\sigma^J\,\sigmab^I\,\omega^L\Big] + \tr\left[\sigmab^I\,\sigma^J\,\omega^L{}^\dagger\right]
\label{isomorphismslLso}
\ee
between the Lie algebras $\sl_L$ and $\so$. Likewise, the homomorphism
\be
\Lambda^I{}_J = \tr\left[\sigmab^I\,\Lambda^R{}^\dagger\,\sigma_J\,\Lambda^R\right]
\ee
between $\SL_R$ and $\SO$ gives an isomorphism
\be
\omega^{IJ} = \tr\left[\sigma^J\,\sigmab^I\,\omega^R{}^\dagger\right] + \tr\,\Big[\sigmab^I\,\sigma^J\,\omega^R{}\Big]
\label{isomorphismslRso}
\ee
between $\sl_R$ and $\so$. On the right-hand side of eqns.\ \eq{isomorphismslLso} and \eq{isomorphismslRso}, the first term is the self-dual part $\omega^+$ of the $\so$--element $\omega$, and the second term is the anti-self-dual part $\omega^-$. This can be seen by observing that
\bea
\frac{1}{2}\,\epsilon^{IJ}{}_{KL}\,\sigma^K\,\sigmab^L = \irm\,\sigma^{[I}\,\sigmab^{J]}\,,
\eea
and therefore
\bea
\omega^+{}^{IJ} &\equiv& \frac{1}{2}\left(\omega^{IJ} - \frac{\irm}{2}\,\epsilon^{IJ}{}_{KL}\,\omega^{KL}\right)
= \tr\,\Big[\sigma^J\,\sigmab^I\,\omega^L\Big]\,, \label{selfdual} \\
\omega^-{}^{IJ} &\equiv& \frac{1}{2}\left(\omega^{IJ} + \frac{\irm}{2}\,\epsilon^{IJ}{}_{KL}\,\omega^{KL}\right)
= \tr\,\Big[\sigmab^I\,\sigma^J\,\omega^R{}\Big]\,. \label{antiselfdual}
\eea
In our model, we take the $\sl_L$ and $\sl_R$ connection to be independent (i.e.\ they do not need to map into the same $\so$ connection), so they give, via eqns.\ \eq{selfdual} and \eq{antiselfdual}, the self-dual and anti-self-dual part of two different $\so$ connections, say, of $A_1$ and $A_2$: 
\bea
A^+_1{}^{IJ} &\equiv& \frac{1}{2}\left(A_1{}^{IJ} - \frac{\irm}{2}\,\epsilon^{IJ}{}_{KL}\,A_1^{KL}\right)
= \tr\,\Big[\sigma^J\,\sigmab^I\,A^L\Big]\,,  \\
A^-_2{}^{IJ} &\equiv& \frac{1}{2}\left(A_2{}^{IJ} + \frac{\irm}{2}\,\epsilon^{IJ}{}_{KL}\,A_2^{KL}\right)
= \tr\,\Big[\sigmab^I\,\sigma^J\,A^R{}\Big]\,.  
\eea
If we plug this into the Lagrangian, we obtain
\be
\clL_{\mathrm{gravity}} = 
\frac{1}{16\pi G}\,\det(E)\,E_I{}^\mu E_J{}^\nu
\left(F_{\mu\nu}{}^{IJ}(A^+_1) + \mathrm{h.c.} + F_{\mu\nu}{}^{IJ}(A^-_2) + \mathrm{h.c.}\right) 
\ee
Variation w.r.t.\ $A_1$ and $A_2$ yields
\be
D^+_{1[\mu} E^I{}_{\nu]} = 0\,, \qquad D^-_{2[\mu} E^I{}_{\nu]} = 0\,.
\ee 
These equations imply that $A^+_1$ is the self-dual part of the spin connection (i.e.\ the connection compatible with the tetrad), and that $A^-_2$ is the anti-self-dual part of the spin connection. By plugging this back into the Lagrangian, we get
\be
\clL_{\mathrm{gravity}} = \frac{1}{8\pi G}\,\det(E)\,E_I{}^\mu E_J{}^\nu
\left(R^+_{\mu\nu}{}^{IJ} + R^-_{\mu\nu}{}^{IJ}\right) 
= \frac{1}{8\pi G}\,\det(E)\,E_I{}^\mu E_J{}^\nu\,R_{\mu\nu}{}^{IJ}\,,
\ee
which is the Einstein-Hilbert action in the tetrad formulation.

This means, in particular, that Minkowski spacetime is a solution, if we only consider the Lagrangian \eq{onlyfirsttwoterms}.

What happens if we include the remaining terms in the gauge Lagrangian?
\bea
\clL_{\mathrm{gauge}} &=& 
\det(E)\left[
\frac{1}{16\pi G}\,E_I{}^\mu E_J{}^\nu\left(\tr\left[\sigma^I\,\sigmab^J F^L_{\mu\nu}\right] + \mathrm{h.c.} \right) 
\right. \\
&& \hspace{1.3cm}{}- \frac{1}{4 g^2}\left(\tr\left[s^{-1}\,F^{L\dagger}_{\mu\nu}\,s\,F^L{}^{\mu\nu}\right]  \right) \\
&& \hspace{1.3cm}{}+ \frac{m^2}{16}\,\tr\left[\left(s^{-1} D_\mu s\right)\left(s^{-1} D^\mu s\right) + \mathrm{h.c.} \right]\\
\eea
Upon variation of $A^L$, $A^L{}^\dagger$ and $s$, we obtain
\be
\frac{1}{8\pi G}\, D^L_\nu\left(E_I{}^\mu E_J{}^\nu\,\sigma^I\,\sigmab^J\right) 
-\frac{1}{2g^2}\, D^L_\nu\left(s^{-1}\,F^{L\dagger}{}^{\mu\nu}\,s\right)
+\frac{m^2}{8}\,s^{-1} D^L{}^\mu s \quad=\quad 0\,,
\label{AL}
\ee
\be
\mathrm{h.c.} \quad=\quad 0\,,
\label{ALhc}
\ee
\be
\frac{1}{4g^2} \left[F^L{}^{\mu\nu} , s^{-1}\,F^{L\dagger}_{\mu\nu}\,s\right] s^{-1} 
- \frac{m^2}{16}\left(D^\mu \left(s^{-1} D_\mu s\right)\right) s^{-1} + \mathrm{h.c.} \quad=\quad 0\,,
\label{sL}
\ee

Thus, we can satisfy all five equations if we can find a solution to
\be
\frac{1}{8\pi G}\, D^L_\nu\left(E_I{}^\mu E_J{}^\nu\,\sigma^I\,\sigmab^J\right) \quad=\quad 0\,,
\ee

\be
\frac{1}{8\pi G}\, D^L_\nu\left(E_I{}^\mu E_J{}^\nu\,\sigma^I\,\sigmab^J\right) \quad=\quad 0\,,
\ee

\be
-\frac{1}{2g^2}\, D^L_\nu\left(s^{-1}\,F^{L\dagger}{}^{\mu\nu}\,s\right)
+ \frac{m^2}{8}\,s^{-1} D^L{}^\mu s \quad=\quad 0\,,
\ee
Clearly, this is solved if we choose tetrads corresponding to the Minkowski metric, $A_L = 0$, $s = \mathrm{const}$.
By a global gauge transformation, we can rotate one of the $s$, say $s$, to $s = \mathds{1}$.

With this configuration, we can also solve the equation of motion arising from variation of the tetrad. 

Thus, we can select any tetrad field $E_I{}^\mu$ corresponding to the Minkoswki metric, $E_{I}^{\mu}= \delta_{I}^{\mu}$, any constant inner product for $s$, and take $(E_I{}^\mu,A_L=0,s)$ as our background.

\section{How does the gravition arise?}

Our unified theory has some new features which distinguish it from ordinary General Relativity and the Electroweak theory by themselves.  First of all, from the perspective of the unified theory, GR and the Electroweak interactions, are both determined by the same chiral gauge group and connection.  Secondly the metricity condition $de=0$ is modified by the presence of the electroweak and the $s$ field.  Therefore, it is important to carry out an analysis to check that we do indeed have a massless spin 2 degree of freedom.

In the previous section we showed that the flat Minkowski background,
$E^I{}_{\mu}E^J{}_{\nu} \eta_{IJ} = \eta{\mu\nu}$, a constant field strenght, $F(A) = Const (i.e. A=0)$ and a constant s all simultaneously solve the field equation.  We regard this as the vacuum solutions of the Isoweak gravity Lagrangian.  However, we want to move over to the Einstein-Hilbert formulation to study the propigation of gravity waves.  This requires us to satisfy the metricity condition for both the background fields which solve all the equations of motion as well as the perturbations of the metricity condition.  When this is satsfied we can use the identity: $E^I{}_{\mu}E^J{}_{\nu} \eta_{IJ} = g_{\mu\nu}$ and work with the Einstein-Hilbert formulation of general relativity.  Let us begin by solving the metricity condition subject the solutions of the equation of motion.

\be
\frac{1}{8\pi G}\, D^L_\nu\left(E_I{}^\mu E_J{}^\nu\,\sigma^I\,\sigmab^J\right) 
=  \frac{1}{2g^2}\, D^L_\nu\left(s^{-1}\,F^{L\dagger}{}^{\mu\nu}\,s\right)
- \frac{m^2}{8}\,s^{-1} D^L{}^\mu s \quad=\quad 0\,,
\label{AL}
\ee

After plugging in the solution of the field equations in vacuum, we obtain:

\be
\frac{1}{8\pi G}\, D^L_\nu\left(E_I{}^\mu E_J{}^\nu\,\sigma^I\,\sigmab^J\right) 
= 0
\label{AL2}
\ee

We now proceed to perturb the metricity conditon to first order so as to find the constraint on the perturbation of the connection.

Focussing on the mass term of the $s$ term we pick out a nonvanishing time-like component of the gauge field.  This is exactly the component that acquires the mass from picking a global 'timelike' direction which breaks the Lorentzian isospin symmetry from $\SL$ to $\SU$.  Therefore, the relevant nonvanishing perturbation in the right hand side of (\ref{AL}) is:

\be
Tr\left[{\twomatrix{1}{0}{0}{1}}\twomatrix{\delta A_{0} + \delta A_{3}}{\delta A_{1} -i\delta A_{2}}{\delta A_{1} +i\delta A_{2}}{\delta A_{0} + \delta A_{3}}_{\mu} {\twomatrix{1}{0}{0}{1}}\right] = 2\delta A^{0}_{\mu} 
\ee
After some algebra, the perturbed condition for the metricity condition is:

\be \frac{1}{8\pi G} \left[ \delta A_{\mu}^{IJ} \wedge \delta^{(\mu}_{J}\delta^{\nu)}_{J}\sigma^{I}\bar{\sigma}^{J} + 2\delta^{(\mu}_{I}\partial_{\nu}e^{\nu)}_{J}\sigma^{[I}\bar{\sigma}^{J]} \right] + \frac{1}{g^{2}} \partial^{\mu}\partial_{[\mu} \delta A_{\nu]} + m^{2} \delta A^{t}_{\nu} = 0 
\ee
where $A^{t}$ is the time-like component of the fluctuation.   This is the component that receives a mass due to the breaking from $\SL$  to $SU(2)$.  The other component of $A_{\mu}^{IJ}$ are the massless modes.  Using the symmetry in the $\mu$, $\nu$ indices we arrive at the final condition for the connection perturbation:
\be\frac{1}{2g^{2}}\partial^{\mu}\partial_{[\mu}\delta A_{\nu]} = - \frac{ m^{2}}\partial A_{\nu} 
\ee
Expanding the fluctuation in Fourier modes $\delta A_{\mu} = \int d^{4}k A(k)e^{ik_{\nu}}x^{\nu}$
we get a modified dispersion realtion for the time-like fluctuation of the connection, which clearly reflects that it is massive:

\be k_{A}^{2} + \frac{m^{2}}{8} = 0 \ee

Furthermore, the other components of the Vector potential will remain a massless spin 1 degree of freedom propigating in Minkowski space-time.  Equipped with this result, we can freely move to the metric variables and treat the solution of the connection fluctuation, as a modification to the stress energy tensor. Therefore, we are left to perturb the Einstein field equations keeping terms up to $O(x^{2})$, where $x$ correspond to the set field variables in the Energy-Momentum tensor.   Since we have solved the metricity conditon and for brevity, we shall now work in the metric variables.

Upon varying the total Lagrangian by the tetrad, and using the identity, $E^I{}_{\mu}E^J{}_{\nu} \eta_{IJ} = g_{\mu\nu}$, we obtain the Einstein field equations:

\be G_{\mu\nu} = \kappa T_{\mu\nu} 
\ee

where 
\be T_{\mu\nu} = c\left[ -F^{\mu\gamma} F^{\nu} _{\gamma} - \frac{1}{4} F_{\gamma \delta} F^{\gamma\delta} + c' (s^{-1}\partial_{mu} s)( s^{-1}\partial_{\nu}s) - g_{\mu\nu} (s^{-1}D_{\alpha} s)( s^{-1}D^{\alpha}s) \right]
 \ee

The linearized wave equation for the transverse-traceless (TT)
gravity wave $\delta g_{\mu \nu} = h_{\mu \nu}$ 
($h_{\mu 0}=0, \nabla_\mu h^{\mu}_{\,\,\nu}=0, h^\mu_{\,\,\mu}=0$) 
can be obtained by perturbing Einstein equations. 
A straightforward calculation gives: 

\be
\Box h_{i}^{\,\,j}(t,\vec{x}) = 
\frac{1}{\sqrt{|g|}}\,\partial_\mu(\sqrt{|g|}\,g^{\mu \nu}\,\partial_\nu)\,
h_{i}^{\,\,j}(t,\vec{x}) =  \kappa( A_{k}^{2})\eta_{i}^{j} + A_{k}^{2} h_{i}^{\,\,j}), \ee

where $A_{k}$ is the mode solution to the perturbation of the vector field due to the perturbed metricity condition.

We have therefore established that the Minkowski background space-time which distinguishes the gravity sector from the electroweak sector, up to first order in perturbation theory, self consistently \footnote{Note that we self consistently perturbed the gauge field only when the gauge coupling was small} provides a graviton moving in the medium of the massive vector field which fills space-time.

\section{Quantum field theory on the background}

We will now quantize our model. Due to the size of the Planck mass, we will treat the gravitational part in the gauge Lagrangian classically, and only quantize
the Yang-Mills part of the gauge Lagrangian plus fermion and Higgs Lagrangian. We will also treat the inner products classically. Quantum fluctuations of $s$ are considered in section \ref{restorationofLorentzinvariance}.

Again, we proceed in a stepwise fashion: to start with we only consider the gauge and fermion Lagrangian, and the Higgs sector will be derived in a future paper \cite{deepak}.
\bea
\lefteqn{\clL_{\mathrm{gauge + fermion}}} \\ 
&=& - \frac{1}{4 g^2}\left(\tr\left[s^{-1}\,F^{L\dagger}_{\mu\nu}\,s\,F^L{}^{\mu\nu}\right] \right) \\
&& \hspace{1.3cm}{}+ \frac{m^2}{16}\,\tr\left[\left(s^{-1} D_\mu s\right)\left(s^{-1} D^\mu s\right)\right] + \mathrm{h.c.} \\
&& \hspace{1.3cm}{}+ \irm\,\psi_L^\dagger\,E_I{}^\mu\,\sigmab^I\,s\,D_\mu\psi_L{} + \mathrm{h.c.} \\
\eea
The tetrad $E_I{}^\mu$ and inner products take their background values, while $A_L$ and the fermion fields are allowed to fluctuate around the background values.

Let us now decompose $A^L$ into parts that are ``anti-hermitian'' and ``hermitian'' w.r.t.\ to $s$:
\[
\renewcommand{\arraystretch}{1.7}
\begin{array}{lcl}
\ds A^L_\mu &=& \ds \At^L_\mu + B^L_\mu\,, \\
\ds \At^L_\mu &=& \ds \frac{1}{2}\left(A^L_\mu - s^{-1} A^L_\mu{}^\dagger s\right)\,, \\
\ds B^L_\mu &=& \ds \frac{1}{2}\left(A^L_\mu + s^{-1} A^L_\mu{}^\dagger s\right)\,, \\
&& \\
 
\end{array}
\]
From this it follows that
\bea
F^L_{\mu\nu} &=& \pa_\mu \At^L_\nu - \pa_\nu \At^L_\mu + [\At^L_\mu,\At^L_\nu] + [B^L_\mu,B^L_\nu]\,, \\
&& {}+ \pa_\mu B^L_\nu + [\At^L_\mu,B^L_\nu] - \pa_\nu B^L_\mu + [\At^L_\nu,B^L_\mu]\,,
\eea
and
\bea
D_\mu s = \pa_\mu s - s\,A^L_\mu - A^L_\mu{}^\dagger\,s = \pa_\mu s - 2 s B^L_\mu\,,\\
\eea
Next we define
\bea
\Ft^L_{\mu\nu} &=& \pa_\mu \At^L_\nu - \pa_\nu \At^L_\mu + [\At^L_\mu,\At^L_\nu]\,, \\
\Dt_\mu B^L_\nu &=& \pa_\mu B^L_\nu + [\At^L_\mu,B^L_\nu]\,.
\eea
Since
\bea
s^{-1} \At^L_\mu s &=& -\At^L_\mu\,, \\
s^{-1} B^L_\mu s &=& B^L_\mu\,,
\eea
we have
\bea
s^{-1}\,F^{L\dagger}_{\mu\nu}\,s 
&=& 
s^{-1}\left[\Ft^L_{\mu\nu}{}^\dagger + [B^L_\mu,B^L_\nu]^\dagger + \left(\Dt_\mu B^L_\nu - \Dt_\nu B^L_\mu\right)^\dagger\right]s \\
&=& 
-\Ft^L_{\mu\nu}{}^\dagger - [B^L_\mu,B^L_\nu]^\dagger + \left(\Dt_\mu B^L_\nu - \Dt_\nu B^L_\mu\right)\,.
\eea
Using all this, the Lagrangian can be expressed in terms of the $\At$ and $B$ components:
\bea
\lefteqn{\clL_{\mathrm{gauge + fermion}}} \\ 
&=& -\frac{1}{4 g^2}\left(
\tr\left[
-\Ft^L_{\mu\nu}\Ft^L{}^{\mu\nu} 
- [B^L_\mu,B^L_\nu]^2 
+ \left(\Dt_\mu B^L_\nu - \Dt_\nu B^L_\mu\right)^2
- \Ft^L_{\mu\nu} [B^L{}^\mu,B{}^L{}^\nu]
\right]\right) \\
&& {}+ \frac{m^2}{16}\,\tr\left[\left(\pa_\mu s - 2 s B^L_\mu\right)^2\right] + \mathrm{h.c.} \\
&& {}+ \left(\irm\,\psi_L^\dagger\,E_I{}^\mu\,\sigmab^I\,s\,D_\mu\psi_L{} + \mathrm{h.c.} \right)\\
\eea
Since the inner products are hermitian, positive definite and of determinant 1, we can always find a gauge transformation that rotates one of them to the identity matrix, say $s = \mathds{1}$. Let us do this and also set
\[
\renewcommand{\arraystretch}{2}
\begin{array}{l@{\qquad}l@{\quad}l}
\nu_L{}^a = \psi_L{}^{a1}\,, & e_L{}^a = \psi_L{}^{a2}\,, \\

\end{array}
\]
Then, the Lagrangian to be quantized is
\bea
\lefteqn{\clL_{\mathrm{gauge + fermion}}} \\ 
&=& -\frac{1}{4 g^2}\left(
\tr\left[
-\Ft^L_{\mu\nu}\Ft^L{}^{\mu\nu} 
- [B^L_\mu,B^L_\nu]^2 
+ \left(\Dt_\mu B^L_\nu - \Dt_\nu B^L_\mu\right)^2
- \Ft^L_{\mu\nu} [B^L{}^\mu,B{}^L{}^\nu]
\right] \right) \\
&& {}+ \frac{m^2}{2}\,\tr\left(B^L\right)^2 + \mathrm{h.c.} \\
&& {}+ \irm\,\nub_L{}^{\ad}\,E_I{}^\mu\,\sigmab^I_{a\ad}
\left[\pa_\mu \nu_L^a + \left(\At^L_\mu{}^a{}_b + B^L_\mu{}^a{}_b\right)\nu_L{}^b + \twomatrix{\At^L_\mu{}^1{}_1}{\At^L_\mu{}^1{}_2}{\At^L_\mu{}^2{}_1}{\At^L_\mu{}^2{}_2} 
\twocolumn{\nu_L{}^a}{e_L{}^a}\right. \\
&& \hspace{2.8cm}{}+ \left.\twomatrix{B^L_\mu{}^1{}_1}{B^L_\mu{}^1{}_2}{B^L_\mu{}^2{}_1}{B^L_\mu{}^2{}_2} \twocolumn{\nu_L{}^a}{e_L{}^a}\right] + \mathrm{h.c.} \\ 
&& {}+ \irm\,\eb_L{}^{\ad}\,E_I{}^\mu\,\sigmab^I_{a\ad}
\left[\pa_\mu e_L^a + \left(\At^L_\mu{}^a{}_b + B^L_\mu{}^a{}_b\right) e_L{}^b + \twomatrix{\At^L_\mu{}^1{}_1}{\At^L_\mu{}^1{}_2}{\At^L_\mu{}^2{}_1}{\At^L_\mu{}^2{}_2} 
\twocolumn{\nu_L{}^a}{e_L{}^a}\right. \\
&& \hspace{2.8cm}{}+ \left.\twomatrix{B^L_\mu{}^1{}_1}{B^L_\mu{}^1{}_2}{B^L_\mu{}^2{}_1}{B^L_\mu{}^2{}_2} \twocolumn{\nu_L{}^a}{e_L{}^a}\right] + \mathrm{h.c.} \\ 
\eea
In this way, we obtain a Lagrangian that contains interaction terms of the standard model as well as non-standard terms.

We choose the mass $m$ sufficiently large, so that the effects of the $B$-particles are unobservable in present accelerators. Thus, we drop all terms involving $B_L$. The remaining minimal coupling terms are of two types: one type is standard model--like, namely,
\be
\irm\,\nub_L{}^{\ad}\,E_I{}^\mu\,\sigmab^I_{a\ad}
\left[\pa_\mu \nu_L^a + \twomatrix{\At^L_\mu{}^1{}_1}{\At^L_\mu{}^1{}_2}{\At^L_\mu{}^2{}_1}{\At^L_\mu{}^2{}_2} \twocolumn{\nu_L{}^a}{e_L{}^a}\right]\,,
\ee
and similarly for $e_L$.

The second type of minimal coupling term is non--standard: if we introduce a basis $\sigma_i/2$, $i=1,2,3$, in the Lie algebra $\mathrm{su(2)}$, we can write them as
\bea
&& \irm\,\nub_L{}^{\ad}\,E_I{}^\mu\,\sigmab^I_{a\ad}\,\At^L_\mu{}^a{}_b\,\nu_L{}^b + \mathrm{h.c.} \\
&=& \irm\,\nub_L{}^{\ad}\,E_I{}^\mu\,\sigmab^I_{a\ad}\,\At^{Li}_\mu (\sigma_i)^a{}_b\,\nu_L{}^b + \mathrm{h.c.} \\
&=& \irm\,\nu_L^\dagger\,E_I{}^\mu\,\sigmab^I \At^{Li}_\mu \sigma_i\,\nu_L + \mathrm{h.c.}\,,
\eea
and similarly for $e_L$.

We find that under a local Lorentz transformation:
\be
\renewcommand{\arraystretch}{1.7}
\begin{array}{lcl}
\psi_L{}^{ab}(x) &\to& \Lambda^a{}_c\,\psi_L{}^{cb}(\Lambda x)\,, \\
\psi_R{}_{\ad \bd}(x) &\to& (\Lambda^{-1})^{\cd}{}_{\ad}\,\psi_R{}_{\cd \bd}(\Lambda x)\,, \\
f^{\mu\ldots}(x) &\to& \Lambda^\mu{}_\nu\, f^{\nu\ldots}(\Lambda x)\,, \qquad\qquad\qquad \mbox{(all other fields)}
\end{array}
\ee
where all spinor indices other than the first index of the fermions remain untransformed and are considered as internal.
We see that the standard model--like terms in the Lagrangian are invariant under this transformation. 
However, the non-standard terms are not invariant under the global Lorentz transformation.  Such interactions are ruled out on large scales.  However, in this model the equivalent of the Higgs mechanism is lacking.  It might be possilble that when the correct mass generation mechanism inherent to this theory is found, these Lorentz violating processes may turn out be to yield predictions for TeV scale physics at the LHC; this issue will be pursued in a future paper.   
\section{Discussion}

The mystery surrounding the physical origin of chirality, weak-isospin and parity violation in the standard Electroweak theory motivated the author to seek a connection with general relativity in a chiral, self-dual formulation.  This was made possilble because the $\SL$ gauge group enjoys a hidden space-time independent $Z_{2}$ 'parity' symmetry which acts on the complex spinors and connections.  As a result,  the unified classical theory enjoys this parity symmetry.  However, when this symmetry is broken by a spontaneous symmetry breaking that chooses a global time like orientation, parity is violated.  Hence the electroweak interactions with parity violation as well as general relativity emerges.   Perturbations around a flat vacuum Minkowski space reveals a propigating spin 2 degree of freedom and a massless vector boson, which are identified as the graviton and weak bosons, respectively.

There is much to be done especially identifying the Higgs mechanism in this model.  We expect that the Higgs field would emerge as a composite degree of freedom, such as a bound state of fermions.  Furthermore, since this mechanism violates Lorentz violation globally, there are non-standard terms in the effective theory which need to confront precision electroweak tests.  We leave this issue for future work.

\section{acknowledments}
While the author was preparing this paper for publication, he noticed the papers by Nesti and Peracci \cite{Nesti:2007ka,Nesti:2007jz} (arxiv:0706.3307) and Nesti arxiv:0706.3304) that implemented a similar mechanism to the one presented in this paper.  While we reach similar conclusions, our mechanism differs in that I only work with one connection variable and also implemented a metric on the internal space to break parity, rendering the gauge group compact.  I want to give special thanks to Florian Conrady for collaborating with me on this project during the course of the past year and for his help with some of the calculations presented in this paper. I would like to give special thanks to BJ Bjorken and Robert Brout for inspiring him to think about parity violation in the Electroweak theory.  I would also like to thank Abhay Ashtekar, Rojesh Pati, Tirthabir Biswas, Lee  Smolin, Deepak Vaid for enlightening discussions.

{99}
\end{document}